\documentclass[a4paper,11pt]{article}
\usepackage[utf8]{inputenc}

\usepackage{titlesec}
\usepackage{float}

\titleformat{\section}[display]
        {\normalfont\small}{}{0pt}{\MakeUppercase}

\titlespacing*{\section}
{0pt}{12pt}{5pt}

\titlespacing*{\subsection}
{0pt}{5pt}{5pt}

\usepackage{setspace}
\usepackage{amsmath}
\usepackage{graphicx}

\doublespace

\title{What season suits you best? Seasonal light changes and cyanobacterial competition.}
\author{\small{G. CASCALLARES}\\ \scriptsize{Consejo Nacional de Investigaciones Científicas y Técnicas} \\ \scriptsize{Centro At\'omico Bariloche, 8400 San Carlos de Bariloche, Rio Negro, Argentina}\\
\\
\small{P. M. GLEISER}\\ \scriptsize{Consejo Nacional de Investigaciones Científicas y Técnicas} \\ \scriptsize{Centro At\'omico Bariloche, 8400 San Carlos de Bariloche, Rio Negro, Argentina}}

\date{}

\begin{document}

\maketitle

Nearly all living organisms, including some bacterial species, exhibit biological processes with a period of about 24 h called circadian (from the Latin circa, about, dies, day) rhythms. These rhythms allow living organisms to anticipate the daily alternation of light and darkness. Experiments carried out in cyanobacteria have shown the adaptive value of circadian clocks. In these
experiments a wild type cyanobacterial strain (with a 24 h circadian rhythm) and a mutant
strain (with a longer or shorter period) grow in competition. In different experiments the external light dark cycle was changed in order to match the circadian period of the different strains, revealing that the strain whose circadian period matches the light-dark has a larger fitness. As a consequence the initial population of one strain grows while the other decays.

These experiments were made under fixed light and dark intervals. However, in Nature
this relationship changes according to the season. Therefore, seasonal
changes in light could affect the results of the competition. Using a theoretical
model we analyze how modulation of light can change the survival of the different cyanobacterial strains. Our results show that there is a clear shift in the competition due to the modulation of light, which could
be verified experimentally.

\section{Introduction}

Circadian rhythms, oscillations with approximately 24 h period in many biological processes, are found in nearly all living organisms. Until the mid-1980s, it was thought that only eukaryotic organisms had a circadian clock, since it was assumed that an endogenous clock with a period of $\tau=24$ h would not be useful to organisms that often divide more rapidly ~\cite{Johnson}. However, in 1985, several research groups discovered that in cyanobacteria there was a daily rhythm of nitrogen fixation ~\cite{Stal,Grobbelaar,Mitsui}. Huang and co-workers were the first to recognize that a strain of Synechococcus, a unicellular cyanobacterium, had circadian rhythms ~\cite{Huang}. This transformed Synechococcus in one of the simplest models for studying the molecular basis of the circadian clock.

The ubiquity of circadian rhythms suggests that they confer an evolutionary advantage. The adaptive functions of biological clocks are divided into two hypotheses. The external advantage hypothesis supposes that circadian clocks allow living organisms to anticipate predictable daily changes, such as light/dark, so they can schedule their biological functions like feeding and reproduction at appropriate times. In contrast to this hypothesis, it has been suggested that circadian clocks confer adaptive benefit to organisms through temporal coordination of their internal physiology (intrinsic advantage) ~\cite{Sharma}. In this case, the circadian clock should be of adaptive value in constant conditions as well as in cyclic environments.

In order to study if circadian clocks provide evolutionary advantages Woelfle and co-workers tested the relative fitness under competition between various strains of cyanobacteria ~\cite{Woelfle}. They carried out experiments where a wild-type strain ($\tau=25$ h) of cyanobacteria and mutant strains, with shorter ($\tau=22$ h) and longer ($\tau=30$ h) periods, were subjected to grow in competition with each other under light-dark (LD) cycles of different periodicity. They found that the strain which won the competition was the one whose free-running period matched closely the period of the LD cycle. This difference in fitness was observed despite the fact that the growth rates were not significantly different when each strain was grown with no competition. Also, mutant strains could outcompete wild-type strains under continuous light (LL) conditions, suggesting that endogenous rhythms are advantageous only in rhythmic environments ~\cite{Woelfle}. This study provided one of the most convincing evidence so far in support of fitness advantages of synchronization between the endogenous period and the period of environmental cycles.

Ouyang et al. suggested an explanation for fitness differences: this could be due to competition for limiting resources or secretion of diffusible factors that inhibit the growth of other cyanobacterial strains ~\cite{Ouyang}. Roussel et al. proposed mathematical models in order to test which of these hypotheses was more plausible ~\cite{Roussel}. They found that the model based on mutual inhibition was consistent with the experimental observations of ~\cite{Ouyang}. In this model the mechanism of competition between cells involves the production of a growth inhibitor, which is produced only during the subjective day (sL) phase, and that growth occurs only in light phase.

Each of the experiments and computational simulations mentioned before had equal amounts of light and dark exposure. However, in Nature the relationship is not constant, and the duration of sunlight in a day changes according to the season and the latitude. The circadian system has to adapt to day length variation in order to have a functional role in optimizing seasonal timing and generating the capacity to survive at different latitudes ~\cite{Hut}.

In this work we will test how day length variation plays a role in the competition between different strains of cyanobacteria. 

\section{The model}

For modelling the growth of each cyanobacterial strain we use the model introduced by Gonze et al. ~\cite{Gonze}, that is based on a diffusible inhibitor with a light sensitive oscillator to represent the cellular circadian oscillator. The evolution equations of cell population $N_i$ and the level of inhibitor $I$ are:

\begin{equation}
\begin{array}{rl}
\displaystyle \frac{dN_i}{dt}=k_iN_i(1-\sum_{j=1}^n N_j), & \displaystyle \frac{dI}{dt}=\sum_{i=1}^n N_i (p_i- \frac{V_{max}I}{K_M+I})
\end{array}
\end{equation}

\begin{equation}
\begin{array}{rl}
k_i=\begin{cases}
      k & \mbox{in L and $i$ in sL or $I<I_c$}\\ 0 & \mbox{otherwise,}
      \end{cases} & p_i=\begin{cases}
      p & \mbox{if $i$ in sL}\\ 0 & \mbox{otherwise.}
      \end{cases}
\end{array}
\end{equation}

In these equations, $N_i$ is the number of cells of strain $i$, $k_i$ is the growth rate of each strain, $p$ measures the rate of inhibitor production, $V_{max}$ is the maximum rate and $K_M$ is the Michaelis constant characterizing the enzymatic degradation of the inhibitor.

Following the work of Gonze et al. we use a modified version of the Van der Pol oscillator to produce sustained circadian oscillations ~\cite{Gonze}. Also, we use the same parameters which were found to be in agreement with experimental data obtained by Woelfle et al. ~\cite{Woelfle}.

\begin{figure}[H]
\centering
\includegraphics[width=8 cm]{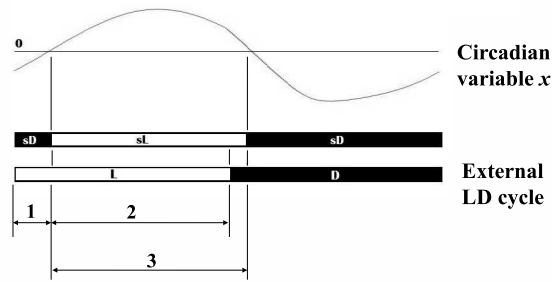}
\caption{Following the schematic explanation of Gonze et al. ~\cite{Gonze}, we show how the model works. The inhibitor $I$ is produced in 3, during the sL phase, and it is degraded during the entire day. Each strain grows in 1, if its sD phase overlaps with L and $I<I_c$, and in 2, when its SL phase overlaps with L.}
\label{fig:model}
\end{figure}

In Fig. \ref{fig:model} we present a schematic plot of the model, that shows how the growth of each population is coupled with the circadian oscillator. As can be seen in this figure, when we modify the length of the light (L) phase the overlaps 1 and 2 change, so the growth of each strain is altered affecting the competition.

\section{Results}

We want to test how day length variation can modify the competition between different cyanobacteria strains. In many organisms, a photoperiodic response is reflected in a physiological change. Photoperiodic responses are common amongst organisms from the equator to high latitudes and have been observed in different types of organisms, from arthropods to plants. Diapause (a suspension of development done by insects), migration and gonadal maturation are examples of these annual changes controlled by photoperiod. These biological processes are triggered as soon as the day length reaches certain duration, known as the critical photoperiod. Even near the equator, where day length changes are very small through the whole year, they are used to synchronize reproductive activities with annual events.

\begin{figure}[H]
\begin{center}
\includegraphics[width=6 cm, angle=270]{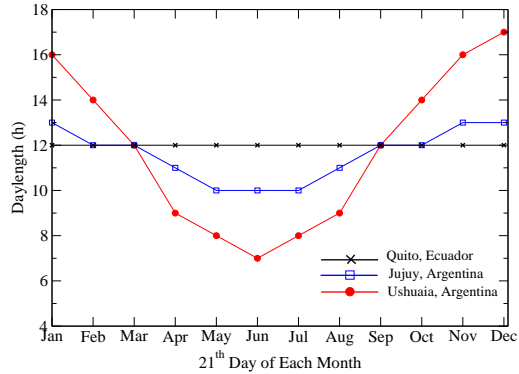}
\end{center}
\caption{Day length over the course of 2012 at different latitudes in South America.}
\label{fig:daylength}
\end{figure}

In Fig. \ref{fig:daylength} we show how day length varies depending on the latitude. The figure compares the day length on the twenty-first day of each month in three cities in South America. We show Quito (Ecuador), which sits near the equator in latitude $0^{\circ}15'$, Jujuy (Argentina), which is located near the tropic of Capricorn in latitude $24^{\circ} 01'$ and Ushuaia, the southernmost city in Argentina which lies in latitude $54^{\circ} 48'$. We can see that during the equinoxes, all places receive 12 hours of daylight.

We simulate the seasonal fluctuations in day length by adding or subtracting minutes of light-time every day to the external LD cycle. For example, if we add 12 minutes of light per day in a LD12:12 cycle, after five days the external cycle of LD has 13 hours of light and 11 of darkness. We initiate competition between equal fractions of wild-type strain and long-period mutant and equal amounts of light and dark. We dilute the culture after 8 light-dark cycles by dividing by a factor of 100 the variables $N_1$, $N_2$ and $I$. In this way we mimic experiments in cultures, that were diluted and sampled every 8 days ~\cite{Woelfle}.

\begin{figure}[H]
\centering
\includegraphics[width=7 cm, angle=270]{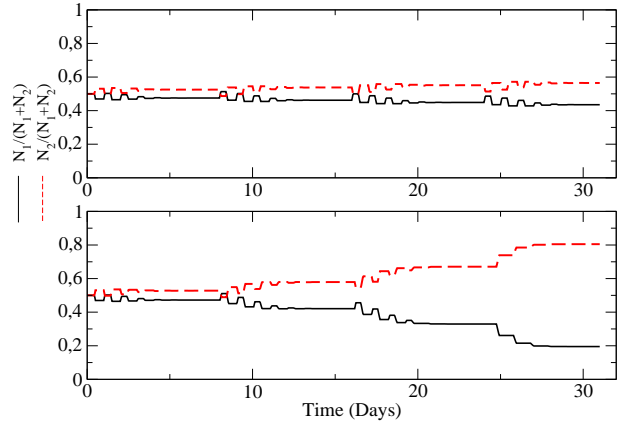}
\caption{(A) The outcome of competition between wild-type (continuous line) and long-period mutant (dashed line) shows coexistence between the two strains for $T=28$ h. (B) Coexistence ends after $\approx 8$ days of adding $3$ minutes of light-time per day. The long-period mutant can win competition as its free-running period becomes closer to the LD cycle.}
\label{fig:fig2}
\end{figure}

First, we analysed the case in which there was a phase of coexistence. For $T=28$, the period of the LD cycle has an intermediate value between the free-running period (FRP) of the two strains and both strains can coexist for a long time. However, when we allowed the days to become longer and the nights shorter, after some days the coexistence was broken, as we show in Fig. \ref{fig:fig2}. This is due to the external period $T$ that starts to get closer to the FRP of the long-period mutant.

\begin{figure}[H]
\centering
\includegraphics[width=9 cm, angle=270]{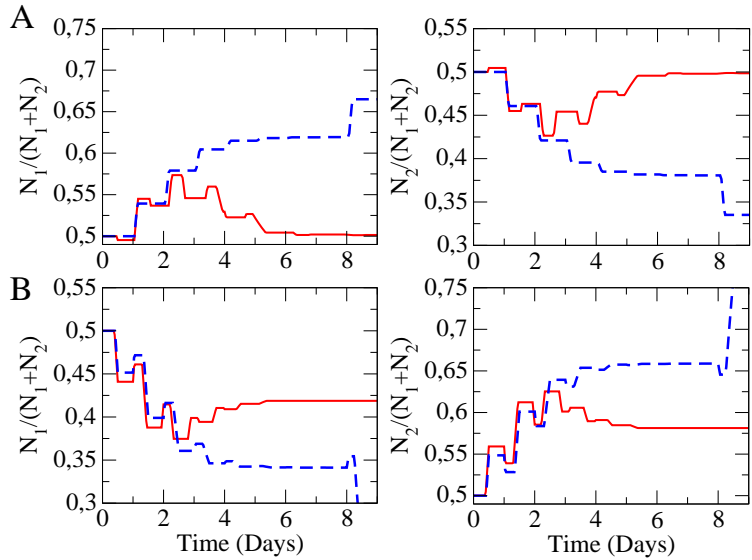}
\caption{Effect of modulation in light time (30 minutes per day) on the outcome of competition between strain 1 (wild-type, $\tau=25 h$) and strain 2 (long-period mutant, $\tau=30 h$) in (A) LD12:12 and (B) LD15:15 cycles. Fraction of strain 1 (left panel) and strain 2 (right panel) are shown as a function of time; red with modulation and blue with fixed cycles.}
\label{fig:n1}
\end{figure}

In Fig. \ref{fig:n1} we show in the left panel the fraction of cells belonging to the wild-type strain as a function of time with fixed LD cycles (red continuous curve) and in the case in which we added 30 minutes of light time each day (blue dashed curve). The first days both curves are similar, but from the second day, the long-period strain has a FRP closest to the period of the LD cycle. The fraction of wild-type strain starts to decrease and is out-competed. In the right panel we show the corresponding fraction of long-period mutant cells in the same cases.

The results of our numerical simulations could be tested experimentally. This would be very simple, since the cultures do not need to be diluted. It is only needed to sample the culture at regular intervals to determine the composition of the population and verify a difference of about ten percent in the two cultures after eight days.

\begin{figure}[H]
\centering
\includegraphics[width=8 cm, angle=270]{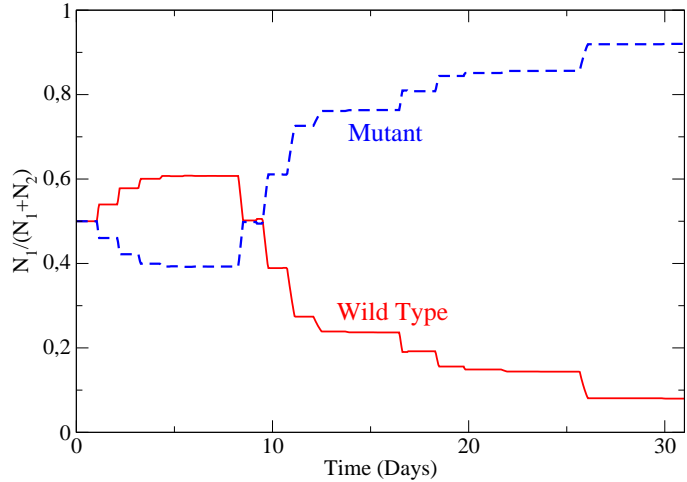}
\caption{Competition between long-period mutant and wild-type strains in a LD12:12 cycle for the same parameters as in Fig. \ref{fig:n1}, but adding 12 minutes per day the light time. A crossover is observed after 8 days.}
\label{fig:fig3}
\end{figure}

Starting from this simple test, we looked for non trivial effects in a longer experiment. We found an interesting effect that can be observed in Fig. \ref{fig:fig3}. In this simulation we added 12 minutes of light time each day. In the first days, the growth was the expected. The wild-type strain could out-compete the long period mutant strain, since the external cycle was LD12:12. But after eight days, when the day was longer than 13 hours, a crossover was observed. The mutant strain started to win the competition because its endogenous period was closer now to the external cycle. This effect could also be tested experimentally, however, in this case the dilution of the culture every 8 days would be needed.

\section{Conclusions}

The mechanisms underlying the enhancement of reproductive fitness remain still unknown. Despite numerous models have been tested, each has some evidence that supports it and none can be excluded at this time ~\cite{JohnsonReview}. In this work we used a diffusible inhibitor model, so our predictions in the growth rates changes could be useful to test the validity of this mechanism. 

Our study is motivated by fluctuations in the day length throughout the year which are reflected in organisms behaviour. We studied how these fluctuations affect the competition between different strains of cyanobacteria. We found non-trivial effects which could be tested experimentally. In the first case we determinate the composition of two strains under competition after eight days when the light is modulated. The prediction of these numerical simulations can be tested in a simple experiment where no dilution is needed. We also propose a second experiment where dilution in the cultures is necessary, which allows for a non trivial effect such as a crossover to be observed.


\begin{thebibliography}{99}

\bibitem{Johnson} Johnson, C.H., S.S. Golden, M. Ishiura, and T. Kondo. Circadian clocks in prokaryotes. \emph{Mol. Microbiol.}
21:5-11 (1996).

\bibitem{Stal} Stal, L.J. and W.E. Krumbein. Nitrogenase activity in the non-heterocystous cyanobacterium Oscillatoria sp. grown under alternating light-dark cycles. \emph{Arch. Microbiol.} 143:67-71 (1985).

\bibitem{Grobbelaar} Grobbelaar, N., T.C. Huang, H.Y. Lin and T.J. Chow. Dinitrogen-fixing endogenous rhythm in Synechococcus RF-1. \emph{FEMS Microbiol. Lett.} 37:173-177 (1986).

\bibitem{Mitsui} Mitsui, A., S. Kumazawa, A. Takahashi, H. Ikemoto, and T. Arai. Strategy by which nitrogen-fixing unicellular cyanobacteria grow photoautotrophically. \emph{Nature} 323:720-722 (1986).

\bibitem{Huang} Huang, T.C., and T.J. Chow. Characterization of the rhythmic nitrogen-fixing activity of Synechococcus RF-1 at the transcription level. \emph{Curr. Microbiol.} 20:23-26 (1990).

\bibitem{Sharma} Vaze, K.M. and V.K. Sharma. On the Adaptive Significance of Circadian Clocks for Their Owners. \emph{Chronobiol. Int.} 30:413-433 (2013).

\bibitem{Woelfle} Woelfle, M.A., Y. Ouyang, K. Phanvijhitsiri and C.H. Johnson. The Adaptative Value of Circadian Clocks: An Experiment Assesment in Cyanobacteria \emph{Curr. Biology} 14:1481-1486 (2004).

\bibitem{Ouyang} Ouyang,Y., C.R. Andersson, T. Kondo, S. Golden and C.H. Johnson. Resonating circadian clocks enhance fitness in cyanobacteria. \emph{PNAS} 95:8660-8664 (1998).

\bibitem{Roussel} Roussel, M., D. Gonze and A. Goldbeter. Modelling the Differential Fitness of Cyanobacterial Strains whose Circadian Oscillators have Different Free-running Periods: Comparing the Mutual Inhibition and Substrate Depletion Hypotheses. \emph{J. Theor. Biol.} 205:321-340 (2000).

\bibitem{Hut} Hut, R.A. and G. M. Beersma. Evolution of time-keeping mechanisms: early emergence and adaptation to photoperiod. \emph{Phil. Trans. R. Soc. B} 366:2141-2154 (2011).

\bibitem{Gonze} Gonze, D., M. Roussel and A. Goldbeter. A Model for the Enhancement of Fitness in Cyanobacteria Based on Resonance of a Circadian Oscillator with the External Light-Dark Cycle. \emph{J. Theor. Biol.} 214:577-597 (2002).

\bibitem{JohnsonReview} Ma, P.,  M. A. Woelfle, and C. H. Johnson. An evolutionary fitness enhancement conferred by the circadian
system in cyanobacteria. \emph{Chaos, Solitons \& Fractals} 50:65-74 (2013).

\end{thebibliography}
\end{document}